\documentclass[prb,
twocolumn,
showpacs,
preprintnumbers,
amssymb, amsmath, floatfix]{revtex4}
\usepackage{graphicx}

\newcommand{\dg}{\ensuremath{\dagger}}
\newcommand{\nn}{\ensuremath{\nonumber}}

\renewcommand{\vec}[1]{\ensuremath{\boldsymbol{\mathrm{#1}}}}

\newcommand{\ket}[1]{\ensuremath{|#1\rangle}}
\newcommand{\bra}[1]{\ensuremath{\langle#1|}}

\begin{document}
\title{Magneto-optical conductivity of Weyl semimetals}%
\author{Phillip E. C. Ashby}
\email{ashbype@mcmaster.ca}
\affiliation{Department of Physics and Astronomy, McMaster University, Hamilton, Ontario, Canada L8S 4M1}

\author{J. P. Carbotte}
\email{carbotte@mcmaster.ca}
\affiliation{Department of Physics and Astronomy, McMaster University, Hamilton, Ontario, Canada L8S 4M1}
\affiliation{The Canadian Institute for Advanced Research, Toronto, Ontario, Canada M5G 1Z8}

\begin{abstract}
Weyl semimetals are a topological phase of matter that have drawn recent interest and have been suggested as a possible phase of the pyrochlore iridates, among other materials.  Here, we compute the magneto-optical response of an isolated Weyl point.  We find that the conductivity is a series of asymmetric peaks lying on top of a linear background.  We find additional features in the conductivity that are explained by the underlying Landau level structure.  We show that weak disorder tends to blur out the peaks, but they still show up as measurable oscillations on a linear background.
\end{abstract}

\maketitle
\section{Introduction}
A Dirac semimetal is a material in which the conduction and valence bands only touch at isolated points within the Brillouin Zone (BZ).  Around these points, the dispersion is linear and the low energy theory is described by a Dirac Hamiltonian. Perhaps the most famous Dirac material is graphene, which contains 2 inequivalent Dirac points in its BZ. These two dimensional Dirac points are not robust against perturbations. Any perturbation proportional to $\sigma_z$ will gap out the band touching. For example, in graphene spin orbit coupling should induce a mass gap.  However, since carbon has such a small atomic number, spin orbit coupling is very weak and this splitting has never been measured in graphene. Furthermore, only intrinsic spin orbit interactions will induce a mass gap. Extrinsic interactions of the Rashba type do not induce a gap in the Dirac spectrum.\cite{Huertas-Hernando:2006fk} 

Dirac semimetals can also exist in three dimensions.  An isolated band touching is described by the Hamiltonian

\begin{align}
\label{eq:WeylH}\mathcal{H} = v_F\vec{\sigma}\cdot\vec{k},
\end{align}
where $v_F$ is the Fermi velocity, $\vec{\sigma}$ is the vector of Pauli matrices, and $\vec{k}$ is the momentum as measured from the band touching. This equation is the electronic analogue of the Weyl equation from particle physics.  Materials described by such a Hamiltonian have become known as \emph{Weyl semimetals}.  The Weyl Hamiltonian is completely robust against any perturbation since it uses all three of the Pauli matrices. In materials with time reversal symmetry ($\mathcal{T}$) and inversion symmetry ($\mathcal{I}$) then there must be four bands linearly dispersing around any band touching point in the BZ. Such a material is called a 3D Dirac semimetal and it is not robust against perturbations since there are additional Dirac matrices in the $4\times 4$ representation.  In a system with broken $\mathcal{T}$ or $\mathcal{I}$ it is possible to have a phase described by the Weyl Hamiltonian.

Weyl points also have interesting topological properties. They are monopoles of Berry flux in the BZ.  This peculiar momentum space topology manifests itself in several ways. One way is the peculiar nature of the gapless surface states. The surface states of a Weyl semimetal are known as Fermi arcs.\cite{Wan:2011fk}  These Fermi arcs are open segments of Fermi surface connecting the projections of the Weyl points onto the 2D surface BZ. These arcs connect Weyl points of opposite chirality and are a signature of the topological nature of the Weyl points. Another manifestation of the topological nature of Weyl points is in the low energy electromagnetic response.  The low energy effective action contains a topological $\theta$-term and is described by axion electrodynamics.\cite{Goswami:2012fk}  The physical effect of this exotic $\theta$-term is an anomalous Hall conductivity.\cite{Burkov:2011kx} The equations of motion also show that these materials could exhibit a chiral magnetic effect -- dissipationless currents in response to an applied magnetic field.\cite{Jian-Hui:2013kx} Recent numerical experiments found no evidence for the chiral magnetic effect using their lattice model for a Weyl semimetal.\cite{Franz:2013fk} Charge transport in Weyl semimetals in the presence of disorder have also been studied, where they found two distinct regimes for the conductivity.\cite{Hosur:2012fk}  The direct observation of Fermi arcs or the anomalous Hall effect would be good evidence for the existence of a Weyl semimetal.

Although Weyl semimetals have not yet been observed there are several candidates.  The pyrochlore iridates \cite{Wan:2011fk,Witczak-Krempa:2012uq,Tafti:2012uq,Ishikawa:2012kx} are a promising host for the Weyl semimetalic phase.  Proposals have also been made for the existence of a Weyl semimetalic phase in topological insulator heterostructures,\cite{Burkov:2011kx,Burkov:2011ys,Zyuzin:2012zr,Halasz:2012ly} as well as in magnetically doped topological insulators.\cite{cho2011:fk,Liu2012:fk}  A recent paper shows that it may even be possible to induce a topological phase transition to a Weyl semimetal using the electron-phonon interaction.\cite{Garate:2013ys}

The optical conductivity in a variety of quasicrystaline materials is linear over a broad range in frequencies.\cite{Timusk:2013fk}  This has been interpreted as evidence for Dirac fermions in these materials.  Further evidence is needed to show that the physics in the quasicrystaline samples is due to Weyl fermions and not simply Dirac fermions.  The magneto-optical response \cite{Fuseya:2011fk, Tabert:2013uq, LiZ:2013gk} is one possible tool that can help distinguish various exotic materials.

In this paper we compute the magneto-optical conductivity for a single Weyl point. For $N_W$ Weyl nodes the conductivity simply scales by a factor of $N_W$.  We find that the transverse optical conductivity is a series of asymmetric peaks with onsets proportional to the square root of the magnetic field. These peaks lie on the background of the zero magnetic field limit.  The structure of these peaks as the chemical potential is changed is rich, and its behaviour follows from the underlying dispersive Landau levels.  We obtain formulae for the absorptive parts of both the transverse and Hall conductivities in the clean limit. Lastly, we examine the conductivity for circularly polarized light, as well as the semi-classical limit. These all provide verifiable predictions distinct to Dirac physics that can give additional support for the existence of a Weyl semimetal.

\section{Dynamical conductivity of a Weyl semimetal}

The Hamiltonian for an isolated Weyl point in a magnetic field is given by
\begin{align}
\mathcal{H} = \left[-i\vec{\nabla}+\frac{e}{c}\vec{A}\right]\cdot\vec{\sigma},
\end{align}
where $e$ is the electron charge, and $c$ is the speed of light. In the above we have ignored the effect of Zeeman splitting, since this effect should be small at accessible magnetic field strengths.  The spin degeneracy is accounted for in the number of Weyl points, $N_W$.   We consider a field $\vec{B} = B\hat{z}$, and use the gauge $A_y=A_z = 0$ and $A_x = -B y$.  We work in units where $\hbar = k_B = v_F = 1$.  It is convenient to define the magnetic length which is given by $l_B^2 = c/eB$. We define the operators
\begin{align}
a = \frac{l_b}{\sqrt{2}}\left(\Pi_x-i\Pi_y\right),\\
a^\dg =  \frac{l_b}{\sqrt{2}}\left(\Pi_x+i\Pi_y\right),
\end{align}
where $\vec{\Pi} = -i\vec{\nabla}+e\vec{A}$. Then, the Hamiltonian is given by the following set of $2\times2$ matrices
\begin{align}
\mathcal{H} = \left(\begin{matrix}
    k_z & \frac{\sqrt{2}}{l_B}a  \\
    \frac{\sqrt{2}}{l_B}a^\dg & -k_z \\
\end{matrix}\right).
\end{align}
First we find solutions to the Eigenvalue problem $\mathcal{H}\psi = E\psi$.  We take $\psi = \left(\begin{matrix}\lambda u_{n\lambda}\\v_{n\lambda}\\\end{matrix}\right)$ and for $n\neq 0$ we obtain
\begin{align}
\pm E_n = \pm\sqrt{\frac{2n}{l_B^2}+k_z^2} = \lambda\sqrt{\frac{2n}{l_B^2}+k_z^2} = E_{n\lambda}.
\end{align}
and
\begin{align}
v_{n\lambda}\frac{\sqrt{2n}}{l_B} = u_{n\lambda}\lambda(E_{n\lambda}-k_z).
\end{align}
\begin{figure}[h]
\centering
  \includegraphics[width=0.8\linewidth]{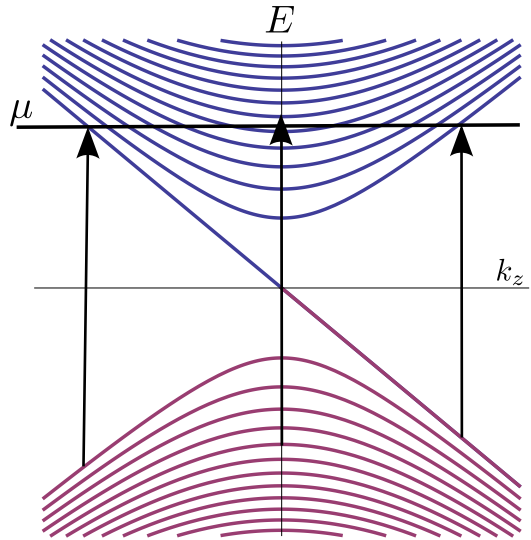}
    \caption{Dispersive Landau level structure for a Weyl semimetal at finite chemical potential. In this figure the chemical potential lies between the fifth and sixth Landau levels. The arrow at $k_z = 0$ shows an interband transition from $n=5$ to $n=6$.  These interband transitions at $k_z=0$ give the main set of peaks shown in Figure \ref{fig:0}. The two transitions at finite $k_z$ occur between the $n=0$ and $n=1$ Landau levels. They occur when $\mu$ lies between two Landau levels and are responsible for the additional bumps seen in Figure \ref{fig:muchange}}.
  \label{fig:landau} 
  \end{figure}

This is the same equation as for a superconductor, with the identification $k_z = \epsilon$, $E_{n\lambda} = E$ and $\sqrt{2n}/l_B = \Delta$.  So
\begin{align}
u_{n\lambda} = \sqrt{\frac{1}{2}\left(1+\frac{k_z}{E_{n\lambda}}\right)},\\
v_{n\lambda} = \sqrt{\frac{1}{2}\left(1-\frac{k_z}{E_{n\lambda}}\right)}.
\end{align}
For $n=0$ we have
\begin{align}
E_0 = -k_z
\end{align}
and
\begin{align}
\psi = \left(\begin{matrix}0\\1\\\end{matrix}\right).
\end{align}

These equations describe the energy level structure and wavefunctions for the set of 3D Landau levels generated by the magnetic field.  Unlike in the 2D case these Landau levels are dispersive (in $k_z$) as shown in Figure \ref{fig:landau}. The structure of these energy levels will control the shape of the dynamical conductivity, as we will elaborate on later.

The dynamical conductivity tensor can be obtained from the Kubo formula. Expressed in the Landau level basis in the clean limit we have
\begin{figure}
\centering
  \includegraphics[width=0.8\linewidth]{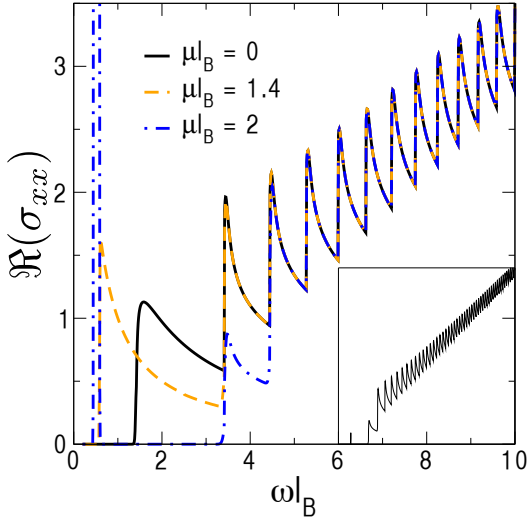}
    \caption{(Color online) The real part of the optical conductivity (in units of $e^2/8\pi l_B$) for $\mu l_B$ = 0,1.4, and 2 as a function of $\omega l_B$. In this figure we took $Tl_B = 0.01$. The conductivity is a series of peaks which sit on a linear background. These peaks occur at the $k_z = 0$ value of the Landau levels shown in Figure \ref{fig:landau}. The inset shows the $\mu$ = 0 conductivity over a larger frequency range, demonstrating its linearity.  The background conductivity should be the result for free fermions, which is a straight line of slope $1/3$ in our units.}
  \label{fig:0} 
  \end{figure}
\begin{widetext}
\begin{align}
\sigma_{\alpha\beta} = -\frac{ie^2}{2\pi l_B^2}\sum_{nn'}\sum_{\lambda\lambda'}\int \frac{dk_z}{2\pi}\left(\frac{f(E_{n\lambda})-f(E_{n'\lambda'})}{E_{n\lambda}-E_{n'\lambda'}}\right)\frac{\bra{\psi_{n,\lambda,k_z}}j_\alpha\ket{\psi_{n',\lambda',k_z}}\bra{\psi_{n',\lambda',k_z}}j_\beta\ket{\psi_{n,\lambda,k_z}}}{\omega+E_{n\lambda}-E_{n'\lambda'}+i0^+},
\end{align}
where $f(x) = 1/(1+e^{\beta(x-\mu)})$, $\beta$ is the inverse temperature, and $\mu$ is the chemical potential. The current operators are given by $j_\alpha = i\left[\mathcal{H},x_\alpha\right] = \sigma_\alpha$.  The chemical potential is related to the density of carriers, $n_0$, through the density of states in a magnetic field, $N(\omega)$, by
\begin{align}
n_0 = \int_{-\infty}^\infty d\omega N(\omega) f(\omega),
\end{align}
\begin{align}
N(\omega) = \frac{1}{2\pi^2l_B^2}\left[1+2\sum_{n=1}^\infty \Re\left(\frac{|\omega|}{\sqrt{\omega^2-2n}}\right)\right].
\end{align} 

We first perform the sum over $\lambda$ and $\lambda'$, this introduces extra terms for the $n=0$ level, but these additional terms all vanish. Then, for the dissipative components of the conductivity tensor $\Re(\sigma_{xx})$ and $\Im(\sigma_{xy})$ we arrive at
\begin{align}
\nn\Re(\sigma_{xx}) = -\frac{e^2}{8 l_B^2}\sum_{n}\int \frac{dk_z}{2\pi}&\left[\left(\frac{f(E_{n})-f(E_{n+1})+f(-E_{n+1})-f(-E_n)}{E_{n}-E_{n+1}}\right)\left(1-\frac{k_z^2}{E_nE_{n+1}}\right)\delta(\omega+E_{n}-E_{n+1})\right.\\
\label{eq:cond1}&+\left.\left(\frac{f(E_{n+1})-f(-E_{n})+f(E_{n})-f(-E_{n+1})}{E_{n}+E_{n+1}}\right)\left(1+\frac{k_z^2}{E_nE_{n+1}}\right)\delta(\omega-E_{n}-E_{n+1})\right],
\end{align}
and
\begin{align}
\nn\Im(\sigma_{xy}) = -\frac{e^2}{8 l_B^2}\sum_{n}\int \frac{dk_z}{2\pi}&\left[\left(\frac{f(E_{n+1})-f(E_{n})+f(-E_{n+1})-f(-E_n)}{E_{n}-E_{n+1}}\right)\left(1-\frac{k_z^2}{E_nE_{n+1}}\right)\delta(\omega+E_{n}-E_{n+1})\right.\\
\label{eq:cond2}&+\left.\left(\frac{f(-E_{n})-f(E_{n+1})+f(E_{n})-f(-E_{n+1})}{E_{n}+E_{n+1}}\right)\left(1+\frac{k_z^2}{E_nE_{n+1}}\right)\delta(\omega-E_{n}-E_{n+1})\right].
\end{align}
\end{widetext}
In this problem the magnetic length, $l_B$ sets a natural energy scale for the problem.  So, for convenience, we define $\bar{\omega} = l_B\omega$, $\bar{k} = l_B k_z$, and $\bar{E}_n = l_B E_n = \sqrt{2n+\bar{k}^2}$. This choice of units will allow us to produce a single plot for all magnetic fields. In the clean limit we can use the delta functions to do the integration over $k_z$. We find that there are contributions from
\begin{align}
\bar{k} = \bar{k}_\pm = \pm\frac{1}{2}\sqrt{\frac{4-4\bar{\omega}^2(2n+1)+\bar{\omega}^4}{\bar{\omega}^2}},
\end{align}
and we obtain the following formulae for the absorptive part of the conductivity tensor:
\begin{widetext}
\begin{align}
\label{eq:final}\Re(\sigma_{xx}) = \frac{e^2}{8\pi l_B}\left[\frac{\sinh(\frac{2+\bar{\omega}^2}{2\bar{\omega}\bar{T}})}{\cosh(\frac{\bar{\mu}}{\bar{T}})+\cosh(\frac{2+\bar{\omega}^2}{2\bar{\omega}\bar{T}})}-\frac{\sinh(\frac{2-\bar{\omega}^2}{2\bar{\omega}\bar{T}})}{\cosh(\frac{\bar{\mu}}{\bar{T}})+\cosh(\frac{2-\bar{\omega}^2}{2\bar{\omega}\bar{T}})}\right]\sum_{n=0}^{\left\lfloor\frac{(\bar{\omega}^2-2)^2}{8\bar{\omega}^2}\right\rfloor}&\left[\frac{\left|2(2n+1)-\bar{\omega}^2\right|}{\bar{\omega}\sqrt{\bar{\omega}^4-4\bar{\omega}^2(2n+1)+4}}\theta\left(|\sqrt{2}-\bar{\omega}|\right)\right],
\end{align}
\begin{align}
-\Im(\sigma_{xy}) = \frac{e^2}{8\pi l_B}\left[\frac{e^{\bar{\mu}/\bar{T}}+\cosh(\frac{2-\bar{\omega}^2}{2\bar{\omega}\bar{T}})}{\cosh(\frac{\bar{\mu}}{\bar{T}})+\cosh(\frac{2-\bar{\omega}^2}{2\bar{\omega}\bar{T}})}-\frac{e^{\bar{\mu}/\bar{T}}+\cosh(\frac{2+\bar{\omega}^2}{2\bar{\omega}\bar{T}})}{\cosh(\frac{\bar{\mu}}{\bar{T}})+\cosh(\frac{2+\bar{\omega}^2}{2\bar{\omega}\bar{T}})}\right]\sum_{n=0}^{\left\lfloor\frac{(\bar{\omega}^2-2)^2}{8\bar{\omega}^2}\right\rfloor}&\left[\frac{|2(2n+1)-\bar{\omega}^2|}{\bar{\omega}\sqrt{\bar{\omega}^4-4\bar{\omega}^2(2n+1)+4}}\theta\left(|\sqrt{2}-\bar{\omega}|\right)\right].
\end{align}
\end{widetext}

\begin{figure}
\centering
  \includegraphics[width=0.8\linewidth]{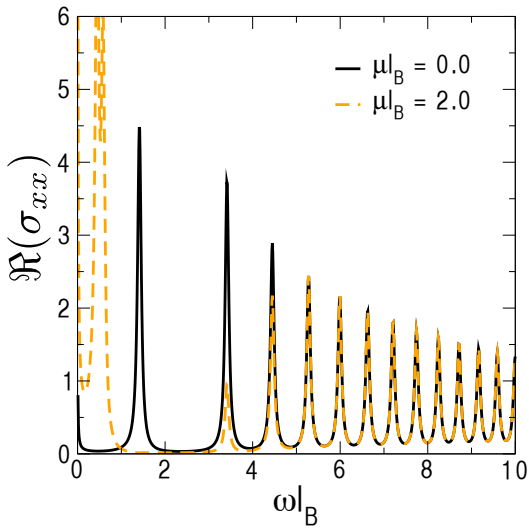}
    \caption{(Color online) The real part of the optical conductivity (in units of $e^2/8\pi l_B$) for graphene in the clean limit. Here $Tl_B = 0.01$. For $\mu=0$ there is a series of peaks at $\bar{\omega}=\sqrt{2(n+1)}+\sqrt{2n}$. As $\mu$ is increased these peaks disappear and are transfered into the intraband peak at low frequency.}
  \label{fig:graphene} 
\end{figure}

In the above expressions $\lfloor x\rfloor$ denotes the integer part of $x$.  $\Re(\sigma_{xx})$ contains a series of peaks that sit on a linear background (Figure \ref{fig:0}).  The background is the result for free Dirac fermions in absence of a magnetic field, namely $\sigma_{0} = \frac{e^2}{24\pi}|\omega|$. In the units of our plots this is a straight line with slope 1/3.  In the inset of Figure \ref{fig:0} we show the real part of $\sigma_{xx}$ over a larger frequency range, showing that it becomes linear at large $\bar{\omega}$ with small oscillations from the Landau level structure.  For $\mu=0$ the peaks occur at $\bar{\omega}=\sqrt{2(n+1)}+\sqrt{2n}$ for integer $n$.  The series of peaks corresponds to allowed interband transitions in the Landau level structure. This peak spacing is proportional to $\sqrt{B}$. From an experimental point of view this means one can see well spaced peaks even for modest fields.  The asymmetry of the peaks is reflected by the square root singularity in Eq. \ref{eq:final}, physically it comes from the dispersive structure of the Landau levels. The long tails of the peaks originate from the square root singularity and add together to provide the linear background we observe.  In Figure \ref{fig:0} we have shown the real conductivity for 3 values of the chemical potential.  These values of the chemical potential are given by the energies of the Landau levels.  We see that as the chemical potential is increased, the peaks at low energies disappear and the optical spectral weight is entirely transfered into an intraband peak at low frequency.
\begin{figure}
\centering
  \includegraphics[width=0.8\linewidth]{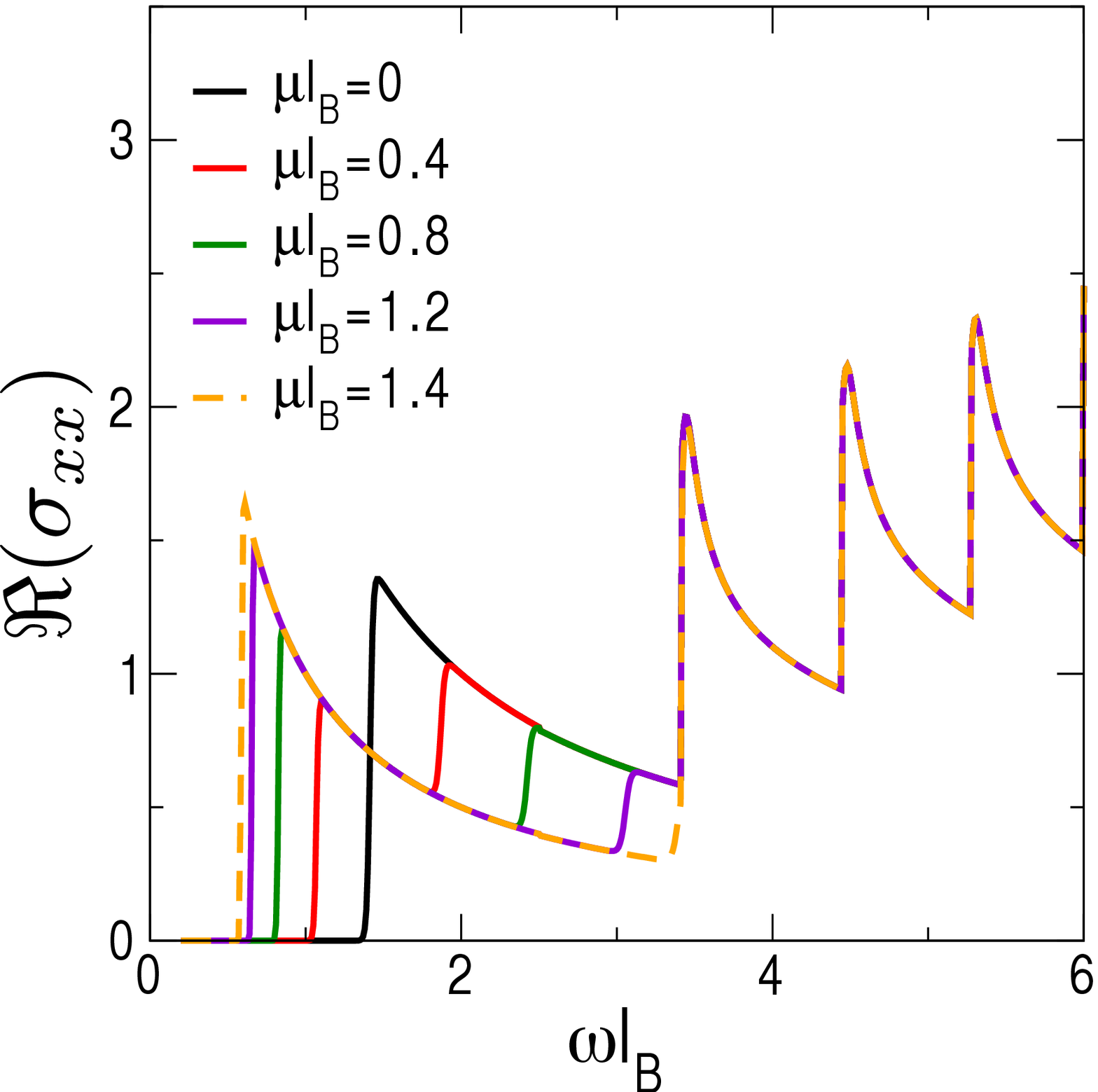}
    \includegraphics[width=0.8\linewidth]{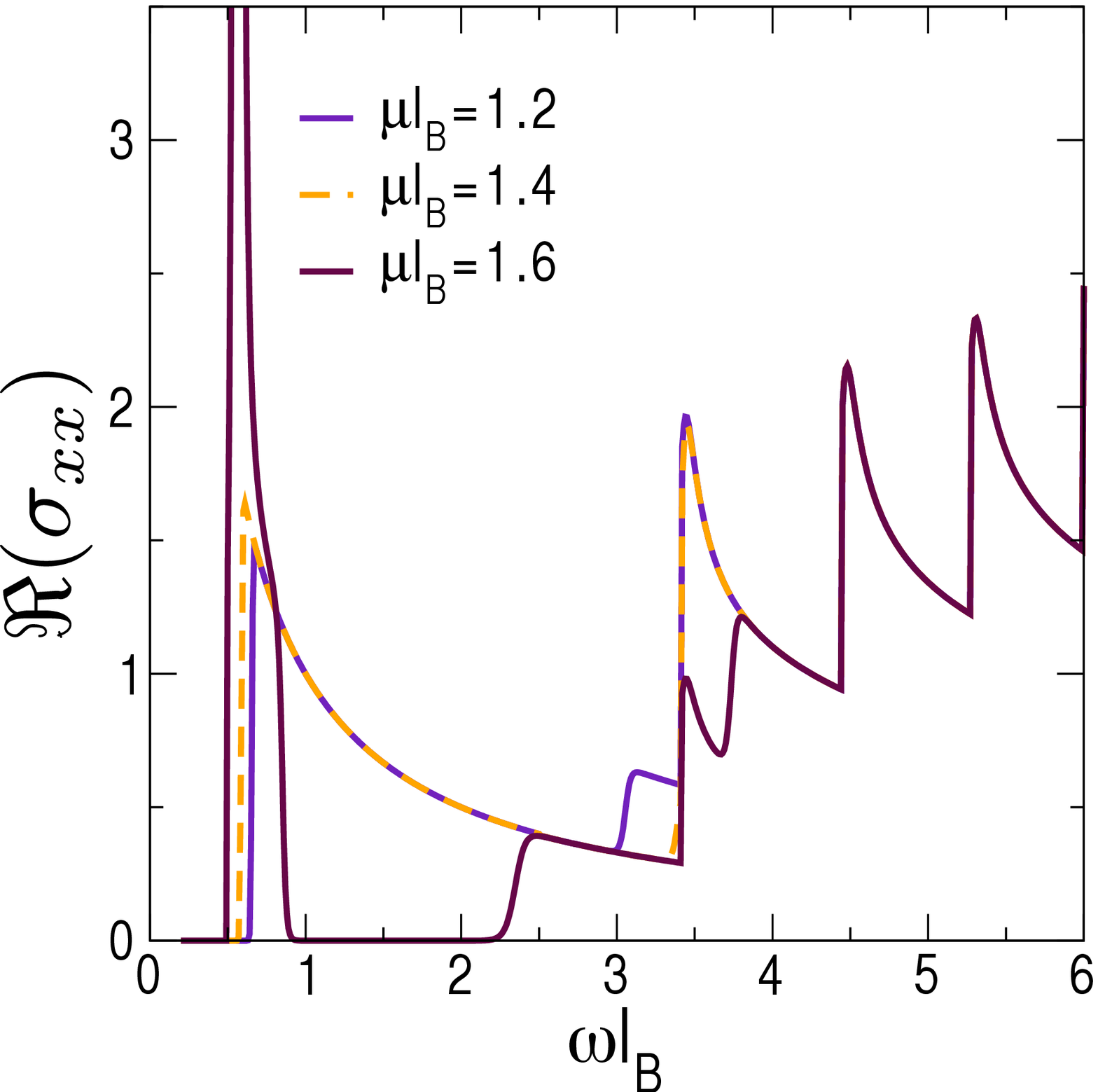}
    \caption{(Color online) The real part of the conductivity(in units of $e^2/8\pi l_B$) as a function of frequency for several values of the chemical potential.  The top panel illustrates how the conductivity changes as the chemical potential sweeps between adjacent Landau levels.  The bottom panel illustrates the change in the optical conductivity as a Landau level is crossed.}
  \label{fig:muchange} 
\end{figure}
The pattern for how the peaks disappear is reminiscent of the what happens in graphene.\cite{Gusynin:2007vn}  For comparison we have calculated the optical conductivity for graphene (Figure \ref{fig:graphene}).  Indeed, the peaks occur at the same positions as for a Weyl semimetal, and disappear as the chemical potential is increased.  However, in graphene the Landau levels are non-dispersive, and changes to the optical conductivity can only happen when the chemical potential passes through a Landau level. In a Weyl semimetal, the dispersive nature of the Landau levels produces a richer structure.  Note the double peak structure associated with the intraband peak in the dasjed curve of Fig. \ref{fig:graphene} at low energy.  One comes from the $1\rightarrow 2$ transtition and the other from the $2\rightarrow 3$ transition.  Both are allowed for $\mu=2$ which falls in the middle of a Landau level.

Figure \ref{fig:muchange} shows how the dispersive Landau level structure affects the optical conductivity.  The top panel shows the evolution as the chemical potential is moved between two adjacent Landau levels.  Unlike in graphene, the spectral weight of the interband peaks is continuously redistributed into the intraband peak.  The bottom panel shows how the conductivity changes as the chemical potential moves through a Landau level. As the chemical potential passes through a Landau level there is a large redistribution of spectral weight, much like in graphene. The lowest peak in the solid black curve disappears entirely as $\mu$ sweeps through the energy of the first Landau level, while the rest of the curve is unaffected. It reappears as a peak in the dashed orange curve that comes from the intraband transition between the $n=1$ and $n=2$ Landau levels.  The tails of this new peak overlap with the original curve.  As $\mu$ is incremented further to 2 (the energy of the $n=2$ Landau level) the second peak in the black curve also becomes modified and now carries only half its original spectral weight.  In a next step the remainder of this second peak will entirely diappear into the intraband transitions. The dispersive Landau levels decorate the graph with additional features. Both the top and bottom panel of Figure \ref{fig:muchange} show additional shoulders that are not located at the signature frequencies $\bar{\omega}=\sqrt{2(n+1)}+\sqrt{2n}$.  These additional features are interband transitions involving the zeroth Landau level. More precisely they are the $n=0\rightarrow n=1$ and $n=1\rightarrow n=0$ transitions. Since the zeroth Landau level is an envelope for all the other Landau levels these additional features exist for all values of the chemical potential.  For a chemical potential $\bar{\mu}$ the shoulders onset at $\bar{\omega}=\bar{\mu}+\sqrt{\bar{\mu}^2+2}$ and $\bar{\omega} = \bar{\mu}+\sqrt{\bar{\mu}^2-2}$.

\begin{figure}
\centering
  \includegraphics[width=0.8\linewidth]{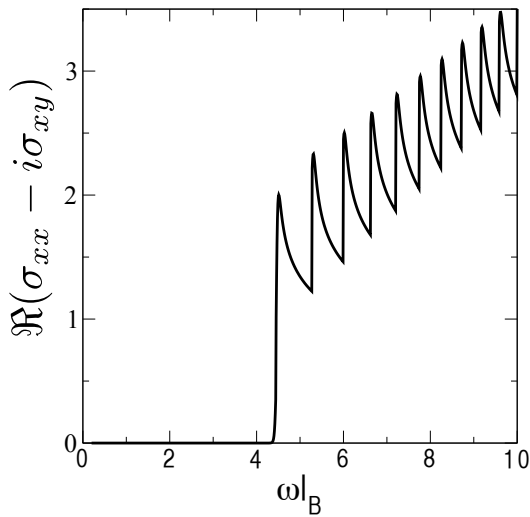}
    \caption{The real part of the optical conductivity (in units of $e^2/8\pi l_B$) for left-handed circularly polarized light for $\bar{\mu} = 2.0$ and $\bar{T} = 0.01$.  In the low temperature limit, the imaginary part of the Hall conductivity exactly cancels the real part of the diagonal conductivity for $\bar{\omega}\lesssim 2\bar{\mu}$.}
  \label{fig:polar} 
\end{figure}
So far we have only discussed the diagonal conductivity $\Re(\sigma_{xx})$.  For experiments that probe the polarization of light, such as the Faraday and Kerr effects, the quantity $\sigma_{\pm} = \sigma_{xx}\pm i\sigma_{xy}$ is appropriate. $\sigma_+$ describes light with right handed polarization, and $\sigma_-$ describes light with left handed polarization.  The absorptive part of $\sigma_{\pm}$ is $\Re(\sigma_{\pm}) = \Re(\sigma_{xx})\mp\Im(\sigma_{xy})$. A plot of $\Re(\sigma_-)$ is shown in Figure \ref{fig:polar}.  For left handed light there is a cancellation between the Hall conductivity and the longitudinal conductivity in the frequency range $\omega\lesssim2\mu$.  For right handed polarization (not shown) the peaks in the region $\omega\lesssim2\mu$ have double the magnitude as in Figure \ref{fig:0}.  The presence of peaks for left circularly polarized light in this region could be a signature of  correlation effects.

\begin{figure}
\centering
  \includegraphics[width=0.8\linewidth]{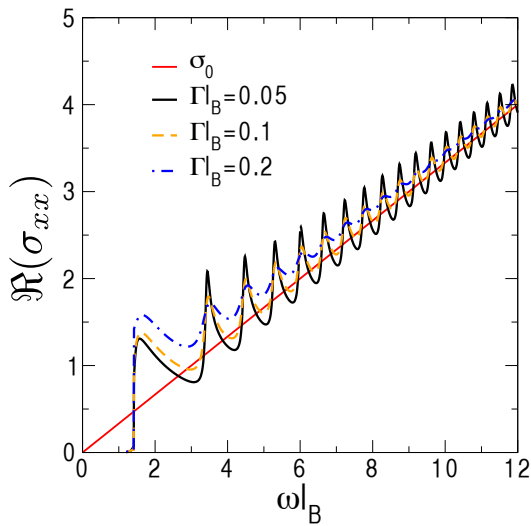}
    \caption{(Color online) The real part of the optical conductivity (in units of $e^2/8\pi l_B$) for $\bar{\mu} = 0.0$ and $\bar{T} = 0.01$ for several values of scattering rate, $\bar{\Gamma}$. The red line shows the conductivity in the absence of a magnetic field.  Increasing disorder tends to blur out the peaks towards the free limit.}
  \label{fig:disorder} 
\end{figure}
\begin{figure}
\centering
  \includegraphics[width=0.8\linewidth]{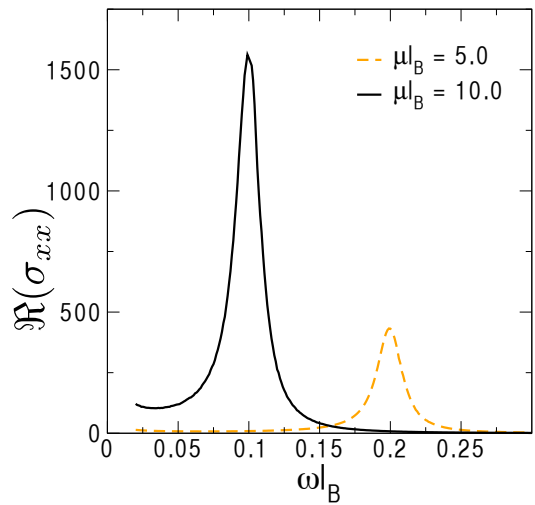}
    \caption{(Color online) The real part of the optical conductivity (in units of $e^2/8\pi l_B$) in the semi-classical limit. This occurs for large values of the chemical potential. All the spectral weight of the now forbidden interband transitions piles up into a single resonance at low frequency.}
  \label{fig:semi} 
\end{figure}

To realistically treat the problem of disorder (especially at large scattering rates) one would need the study of the Kubo formula expressed in terms of the spectral density, $A(k,\omega)$. By solving a Boltzmann equation, one could obtain an expression for the self energy.  The imaginary part of the self energy then serves to broaden the spectral densities.  Here we capture the effect phenomenologically by introducing a scattering rate, $\Gamma$. By making the replacement
\begin{align}
\delta(z)\rightarrow \frac{1}{\pi}\frac{\Gamma}{\Gamma^2+z^2},
\end{align}
in Eqn \ref{eq:cond1} we can understand how weak scattering will effect the conductivity.  This replacement should give a qualitatively correct picture for small values of the inverse scattering time, $\bar{\Gamma}$. 

Figure \ref{fig:disorder} shows the real part of the transverse conductivity for $\bar{\Gamma} =$ 0.05, 0.1, and 0.2.  The straight red line shows the conductivity of free 3D Fermions, $\sigma_0$ for comparison. The disorder tends to smear out the peaks, and at large $\omega l_B$ the conductivity trends towards the free limit as the disorder is increased. In particular the scattering rate needs to be large compared to the energy level spacing to see any peaks.  In Figure \ref{fig:disorder} the first few peaks are still very well defined for $\bar{\Gamma} = 0.2$.  For a Fermi velocity of $v_F\sim 10^6 m/s$, and $B$ measured in Tesla, this corresponds to a scattering rate of $\Gamma \sim 7\sqrt{B} meV$.  

Finally we consider the semi-classical limit.  The semi-classical limit occurs when the Landau level quantization is no longer important.\cite{Pound:2012vn}  This occurs when the chemical potential is the largest energy scale in the problem.  For $\bar{E}_N<\bar{\mu}<\bar{E}_{N+1}$, with $N\gg 1$ we have that
\begin{align}
\Delta\bar{E} = \bar{E}_{N+1}-\bar{E}_N \simeq \frac{1}{\bar{\mu}}.
\end{align}
In this limit, if one goes back to the unbarred units, the energy spacing goes like $B$ in contrast to the $\sqrt{B}$ behaviour at small $\mu$.  The semi-classical resonance it shown in Figure \ref{fig:semi} for $\bar{\mu} =$ 5, and 10.  It appears as a peak at $1/\bar{\mu}$, and contains the spectral weight from the interband transitions that have disappeared between $0<\bar{\omega}<2\bar{\mu}$.  In graphene where the Landau levels are flat this line is due only to the $E_N\rightarrow E_{N+1}$ intraband transition.  In the present case however, the dispersive Landau level structure gives contributions to this lineshape from the set of transitions $E_n\rightarrow E_{n+1}$ with $n\in[0,N]$.  We can also see in Figure \ref{fig:semi} that as we doubled $\mu$ the amount of area in the semi-classical peak quadrupled.  This is a reflection of the linear background of the conductivity. For a chemical potential of $\mu$, interband transitions are forbidden out to $\omega=2\mu$.  The linearity of the conductivity combined with a sum rule shows that the area under this peak should scale like $\mu^2$, as we observe.

\section{Discussion and Conclusions}
In this paper we calculated the optical conductivity for a Weyl semimetal in the presence of a magnetic field. We found that the dispersive Landau level structure produced a set of asymmetric peaks on a linear background. The peaks have a $\sqrt{B}$ spacing, a signature of Dirac physics. The shape of the peaks reflects a square root singularity, and their long tails conspire together to give the linear background in this case. The positions of the peaks occur at the same frequencies as in graphene, although there are additional features when the chemical potential falls between two Landau levels. We also showed that for left hand polarized light, that there is a cancellation for frequencies $\omega<2\mu$, and peaks in this region could be associated with interaction effects. We showed that the semi-classical resonance has a typical lineshape, but that it consists of many intraband transitions and has spectral weight thats scale like $\mu^2$.  We also showed the effect of weak disorder.  The sharp peaks tend to blur out towards the linear background as the scattering rate increased. This may make the peaks hard to see in dirty samples. In particular, the scatting rate needs to be less than the Landau level spacing for the peaks to still be observable.

Finally we would like to make a remark on units.  Throughout the paper we have used the dimensionless units where $l_B$ set the energy scale.  For convenience we include the conversion to `real' units. For $v_F$ measured in $m/s$ and $B$ measured in $T$ we have
\begin{align}
l_B^{-1} = 36.3 v_F\sqrt{B} \times 10^{-9} eV.
\end{align}
This conversion will be useful for comparing to experimental results.  For the quasicrystals the chemical potential sits at the Fermi level. The estimated value for the Fermi velocity is $v_F = 4.3\times 10^7 cm/s$.\cite{Timusk:2013fk}  This sets the magnetic energy at $l_B^{-1} = 15.6\sqrt{B} meV$ for the quasicrystaline candidates.  In the pyrochlore iridates (such as Y$_2$Ir$_2$O$_7$) the chemical potential also sits at the Fermi level and the Fermi velocities are estimated to be about an order of magnitude smaller than in graphene.\cite{Wan:2011fk}  In this case the magnetic energy is approximately $l_B^{-1} = 3.63\sqrt{B} meV$.  At accessible magnetic fields $0<\sqrt{B}\lesssim 6$, and so the effects presented in this paper should be observable at reasonable energies.
\begin{acknowledgements}
This work was supported by the Natural Sciences and Engineering Research Council of Canada and the Canadian Institute for Advanced Research. We thank Tom Timusk for encouraging us to pursue this topic.
\end{acknowledgements}

\bibliography{biblio}

\end{document}